\documentclass[12pt]{iopart}
\usepackage{graphicx}
\usepackage{iopams}

\begin{document}

\title[Affleck-Dine baryogenesis in anomaly-mediated SUSY breaking]
{Affleck-Dine baryogenesis in anomaly-mediated SUSY breaking}
\author{Masahiro Kawasaki, Kazunori Nakayama}
\address{Institute for Cosmic Ray Research,
The University of Tokyo,
Kashiwa 277-8582, Japan}

\begin{abstract}
It has been known that in anomaly-mediated SUSY breaking model
Affleck-Dine baryogenesis does not work
due to trapping of Affleck-Dine field into charge-breaking minima.
We show that when finite-temperature effect is properly taken into account 
and if reheating temperature is relatively high,
the problem of falling into charge breaking global minima can be avoided
and hence Affleck-Dine baryogenesis works.
Moreover, for the $LH_u$ flat direction we obtain a 
constraint on the mass of neutrino. 
\end{abstract}



\section{Introduction}

The origin of baryon asymmetry, or matter-anti-matter asymmetry, is one of the 
most interesting topics for both cosmology and particle physics.
From WMAP three year results~\cite{Spergel:2006},
\begin{equation}
	 \eta=\frac{n_B}{n_{\gamma}} \simeq (6.1 \pm 0.2)
	  \times 10^{-10} .
\end{equation}
is obtained. 
Primordial big-bang nucleosynthesis explains the observed light element abundances
for about the same value of $\eta$~\cite{Olive:2000}. 
To explain this value, various baryogenesis mechanism have been considered.

Supersymmetry (SUSY)~\cite{Nilles:1984} is the most attractive 
candidate as physics beyond the standard model.
Thus it is worthwhile to construct the baryogenesis model 
based on SUSY. Affleck-Dine mechanism~\cite{Affleck:1985} 
is one of the most studied baryogenesis scenario 
in the framework of supersymmetric standard model~\cite{Enqvist:2003}. 
This uses the dynamics of flat directions existing in 
the minimal supersymmetric standard model (MSSM), 
which is constructed of the scalar fields having flat potential 
in supersymmetric limit at renormalizable level. 
The angular motion of flat directions $\phi$ in complex plane
generates lepton or baryon current,
\begin{equation}
	n=iN(\dot \phi^* \phi - \phi^*\dot \phi)=N|\phi |^2 
	\dot \theta  \label{charge}
\end{equation}
where $N$ is the constant determined by the flat direction, 
and we have defined $\phi=|\phi |e^{i\theta}$.
How this angular motion is generated or what amount of 
baryon asymmetry can be created depends on the type of 
flat direction, temperature of the universe and 
the mechanism of supersymmetry breaking.
Finite-temperature effects~\cite{Allahverdi:2000,Anisimov:2001} 
are another important correction,
which can significantly affect the whole dynamics.

Furthermore, formation of Q-balls makes the usual Affleck-Dine 
scenario complicated~\cite{Coleman:1985,Kusenko:1997}.
The property of Q-balls strongly depends on SUSY breaking 
mechanism.
In gravity-mediation scenario, Q-ball is unstable against 
decay into nucleons and created baryon number is finally 
converted to ordinary matter~\cite{Enqvist:1998,Kasuya:2000}.
In gauge-mediation scenario, however, Q-ball is stable and only
evaporation and diffusion effects can extract the baryon 
number from Q-balls~\cite{Kusenko:1998,Laine:1998,Banerjee:2000}.
On the other hand, Q-ball can explain naturally why the dark 
matter and baryon number density are roughly the same order 
in the universe in some specific models~\cite{Fujii&Yanagida:2002}.
Therefore, when considering Affleck-Dine baryogenesis scenario, 
one must carefully trace the dynamics of flat directions 
taking into account various effects.

Anomaly-mediation models~\cite{Randall:1999} have 
attractive feature on phenomenological point of view.
In this scenario, SUSY breaking effect in hidden sector 
are transmitted to observable sector through 
super-Weyl anomaly.
This predicts model-independent generic feature at 
low energy physics insensitive to physics at high energy scale.
As a consequence, the flavor problem existing in usual 
gravity-mediation scenario is relaxed, and the new possibility 
of wino-like dark matter is provided~\cite{Moroi:2000,Fujii:2002}.
Gravitino mass becomes two orders of magnitude larger than 
that of gravity-mediation case, which also solves 
the gravitino problem.
However, as explained in the next section, 
Affleck-Dine baryogenesis in anomaly-mediation models has 
been revealed to be difficult~\cite{Kawasaki:2001,Fujii:2002}.

In this paper, we study Affleck-Dine mechanism in 
anomaly-mediation models including finite-temperature effects,
which was missed in previous literature.
It is shown that when finite-temperature effect is properly 
taken into account  Affleck-Dine baryogenesis works.
We also discuss Q-ball formation and its consequences.

This paper is organized as follows.
First in Sec.~\ref{sec:potential}, potentials for 
flat directions including finite-temperature effects
are given.
In Sec.~\ref{sec:n=4} we describe the dynamics of $n=4$ 
flat directions when finite-temperature effects are 
taken into account.
The case with $n=6$ flat direction is discussed in 
sec.~\ref{sec:n=6}.
In Sec.~\ref{sec:Q-ball} the effects of Q-ball formation 
is described and we conclude in Sec.~\ref{sec:conclusion}

\section{Potential of Affleck-Dine field 
in anomaly-mediation model}   \label{sec:potential}

\subsection{Zero-temperature potential}

First, we summarize the standard scenario for Affleck-Dine 
baryogenesis neglecting the finite-temperature effect.
Potentials of flat directions in the MSSM are exactly flat in 
renormalizable and supersymmetric limit, but lifted by 
non-renormalizable terms and supersymmetry breaking effect.
If we parameterize a flat direction $\phi$, non-renormalizable superpotential
\begin{equation}
	W= \frac{\phi ^{n}}{nM^{n-3}}
\end{equation}
generates the potential for $\phi$,
\begin{equation}
	V= \frac{|\phi |^{2(n-1)}}{M^{2n-6}}
\end{equation}
where $M$ denotes some cut-off scale
\footnote{We use the same symbol $\phi$ as a chiral 
superfield or its scalar part.}.
All MSSM flat directions are known to be lifted up 
to $n=9$ gauge-invariant superpotentials.
Including supersymmetry breaking effects, 
the potential would be 
\begin{eqnarray}
	V = (m_{\phi}^2-cH^2)|\phi |^2   
	+\left\{a_m\frac{m_{3/2}\phi^n}{nM^{n-3}} 
	+a_H\frac{H\phi^n}{nM^{n-3}} +\mathrm{h.c.}\right\}
	+ \frac{|\phi |^{2(n-1)}}{M^{2n-6}},  
	\label{potential}
\end{eqnarray}
where $H$ is Hubble parameter, $c, a_m$ and $a_H$ are 
constants of $O(1)$.
Initially, the $\phi$ field is trapped at the minimum 
determined by the negative Hubble mass term $-H^2|\phi |^2$ 
and the highest term $|\phi |^{2(n-1)}/M^{2n-6}$ as 
\begin{equation}
	|\phi | \simeq (HM^{n-3})^{\frac{1}{n-2}}. 
	\label{minimum}
\end{equation}
The $\phi$ field traces this minimum until $H$ becomes less 
than $m_{\phi}$ and it begins to oscillate around the origin.
The angular direction of $\phi$ is determined by 
the Hubble-induced A-term until this epoch.
If $m_\phi \sim m_{3/2}$ as expected in gravity mediation,
at the same time when $\phi$ begins to oscillate, 
the soft A-term begins to dominate over the Hubble-induced 
A-term, and this causes the kick in the angular direction.
Finally the $\phi$ field shows the $U(1)$ conserving 
elliptical motion around their minimum.
At this stage, the created baryon number is conserved.
This is the standard scenario for creating baryon 
asymmetry in the Affleck-Dine mechanism.

In anomaly-mediation models, however, there is a subtlety.
Since soft mass is loop-suppressed in anomaly-mediation,
we expect $m_\phi \sim m_{3/2}/(8\pi^2)$.
If we assume soft masses should be $100\sim 1000$~GeV, 
the natural order of gravitino mass is estimated to be 
$10 \sim 100$ TeV, two orders of magnitude larger than 
that of gravity-mediation case.
This is a good feature for avoiding gravitino problem.
For such large gravitino mass, its lifetime naturally 
becomes shorter than 1~sec, which does not much affect BBN.
In fact, even if its hadronic branching ratio is order one, 
there is no upper bound on the reheating temperature
if gravitino mass is as large as 100TeV~\cite{Kawasaki:2004yh}.  
However, this invalidates the usual Affleck-Dine 
baryogenesis scenario because the potential of flat 
direction (\ref{potential}) has charge and/or color breaking 
global minima with 
\begin{equation}
	|\phi |_{\mathrm{min}}\sim 
	\left(\frac{|a_m|}{n-1} m_{3/2}M^{n-3} \right)
	^{\frac{1}{n-2}}. 
	\label{breakmin}
\end{equation}
The minimum value of the potential becomes
\begin{equation}
	V(|\phi |_{\mathrm{min}}) \sim  
	-\frac{n-2}{n}M^4 
	 \left( \frac{|a_m|}{n-1}\frac{m_{3/2}}{M} 
	 \right)^{\frac{2n-2}{n-2}},
	 \label{potmin}
\end{equation}
which is always negative for $n\geq 4$.
This is not a problem if the relevant fields sit 
at the origin initially, since the decay rate of the false 
vacuum into a true charge breaking minimum is sufficiently 
small and the lifetime is longer than the age of 
the universe~\cite{Kawasaki:2001}.
But in Affleck-Dine set-up, the corresponding flat direction 
should have large field value 
tracking  their minimum (\ref{minimum}), and 
finally fall into charge breaking minima (\ref{breakmin}).
This is a fundamental problem when applying Affleck-Dine mechanism 
to anomaly-mediation models.

There is an attempt for Affleck-Dine baryogenesis in 
anomaly-mediation models based on gauged $U(1)_{B-L}$ 
symmetry~\cite{Fujii:2001no}. 
This uses the fact that $U(1)_{B-L}$ breaking effect stops 
the $\phi$ field moving beyond the $U(1)_{B-L}$ breaking 
scale $v$ due to $D$-term contribution.
If $v$ is smaller than the field value corresponding to 
the hill of the potential~(\ref{potential}),
\begin{equation}
	|\phi |_{\mathrm{hill}}\sim 
	\left( \frac{m_\phi^2 M^{n-3}}{m_{3/2}} 
	\right)^{\frac{1}{n-2}},
\end{equation}
we do not need to worry about falling into the unphysical 
global minima.
This is an appealing feature, but this model relies on 
the hypothesis of gauged $U(1)_{B-L}$ symmetry.
In the following, we show that finite-temperature effect 
enables us to obtain baryon asymmetry and avoid the charge 
breaking minima in anomaly-mediation models without any 
further assumption beyond MSSM.

\subsection{Finite-temperature effect}

Thermal effects modify the potential of flat directions.
First, couplings of flat directions with other particles 
$\psi$ yields thermal mass term~\cite{Allahverdi:2000} 
given by
\begin{equation}
	\sum _{f_k|\phi |<T}c_k f_k^2 T^2|\phi |^2,
\end{equation}
where $c_k$ is a constant of $O(1)$ and $f_k$ denotes gauge or 
Yukawa coupling relevant for the flat direction.
Note that when $f_k|\phi |>T$, $\psi$ receives a large mass 
of order $f_k|\phi |$ and can not be thermalized, and hence
$\phi$ does not feel thermal mass.

It was also pointed out that the following form of the 
potential~\cite{Anisimov:2001}
\begin{equation}
	V\sim a\alpha(T)^2 T^4 
	\log \left(  \frac{|\phi |^2}{T^2}  \right)
	\label{eq:thermal_log}
\end{equation}
should be included for the potential of flat directions,
where $a$ is a order 1 constant determined by 
two-loop finite temperature effective potential.
For $LH_u$ direction, $a=9/8$.
When this term dominates, it is possible that flat direction 
begins to oscillate due to thermal logarithmic term.
The epoch of onset of oscillation is determined by
\begin{equation}
	H_{\mathrm{OS}}^2\sim m_\phi^2 + 
	\sum _{f_k|\phi |<T}c_k f_k^2 T^2 
	+ a\alpha(T)^2 \frac{T^4}{|\phi |^2}.
	\label{osc.condition}
\end{equation}
Details of the dynamics depend on corresponding flat directions
and somewhat complicated~\cite{Asaka:2000,Fujii:2001}.
Now let us investigate it for $n=4$ and $n=6$ case.

\section{The case of $n=4$ flat direction } \label{sec:n=4}

In this section, we describe the dynamics of AD field for $n=4$
based on the potential given in the previous section,
concentrating on $LH_u$ direction particularly.
This is because Affleck-Dine baryogenesis can not work 
successfully for other $n=4$ directions
(see Sec.\ref{sec:comments}).

\subsection{Dynamics of $n=4$ flat direction}

From eq.(\ref{osc.condition})
for $n=4$, early oscillation due to the thermal log term occurs when
\begin{equation}
	T_R ~\gtrsim ~ \frac{m_\phi}{\alpha(T)}
	\left( \frac{M}{M_P} \right)^{\frac{1}{2}},
\end{equation}
where $T_R$ is reheating temperature after inflation, and
in this case the Hubble parameter at start of the oscillation 
is 
\begin{equation}
	H_{\mathrm{OS}} ~\sim ~ \alpha T_R 
	\left( \frac{M_P}{M} \right)^{\frac{1}{2}}.  
	\label{Hos}
\end{equation}
Here we have used $T\sim (T_R^2 HM_P)^{1/4}$ and 
$|\phi |\sim (HM)^{1/2}$.
For natural range of cut-off scale $M$, early oscillation 
naturally takes place unless reheating temperature $T_R$ 
is unnaturally low.

It should be noticed that, for sufficiently high reheating 
temperature, this thermal logarithmic potential can hide 
the unwanted valley of the potential.
Substituting $n=4$ into eq.(\ref{potmin}),
the minimum of the zero-temperature potential is given by 
\begin{equation}
	V(|\phi |_{\mathrm{min}})\sim -\frac{1}{54}m_{3/2}^3M.
\end{equation}
In order to avoid falling into this minimum, 
at least $\alpha^2 T^4 \gtrsim |V(|\phi|_{\mathrm{min}})|$ 
is required at the beginning of oscillation.
This leads to 
\begin{equation}
	T_R ~\gtrsim ~\alpha^{-1} m_{3/2}
	\left( \frac{M}{M_P} \right)^{\frac{1}{2}}. \label{TRcondition}
\end{equation}
If we assume $M\sim M_P$, $T_R\gtrsim 10 m_{3/2} 
\sim 10^{5-6}$~GeV is needed to satisfy the above 
condition.
Note that thermal mass term can not dominate over the thermal
logarithmic term at this epoch.
We have checked numerically that 
the above condition is almost sufficient 
to drive the Affleck-Dine field into the origin
\footnote{
As an another condition for avoiding global minima,
one may require
$|\phi |<|\phi |_{\mathrm{hill}}$ at $H\lesssim m_{3/2}$,
which is the epoch soft A-term begins to dominate over 
the Hubble-induced A-term.
This is achieved when the following condition is satisfied,
\begin{equation}
	T_R \gtrsim \alpha^{-1} \frac{m_{3/2}^2}{m_\phi} 
	\left( \frac{M}{M_P}\right)^{\frac{1}{2}}.
\end{equation}
But in fact this condition is too strong.
This condition is sufficient, but not always necessary.
Numerical calculation shows the condition (\ref{TRcondition}) is almost sufficient. 
}.

\begin{figure}[htbp]
	\begin{center}
		\includegraphics[width=0.7\linewidth]{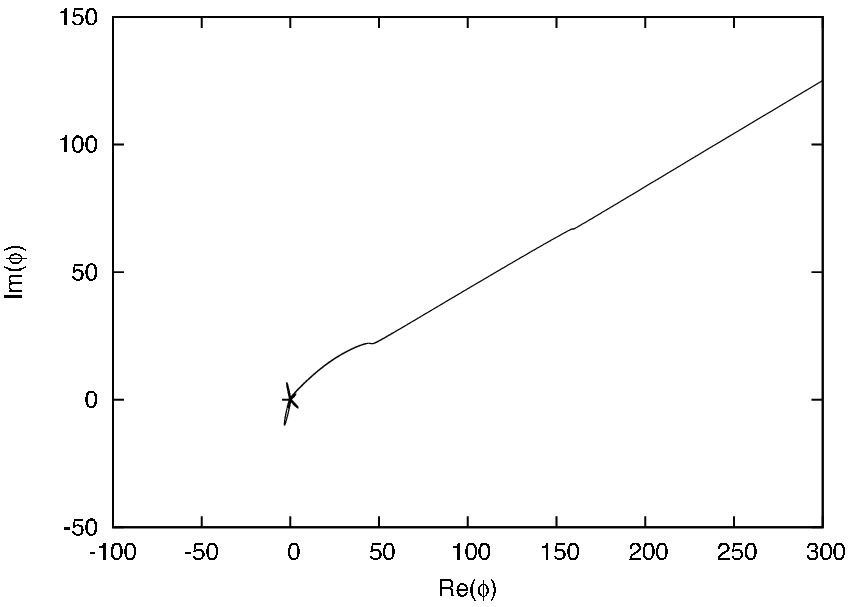}
		\includegraphics[width=0.7\linewidth]{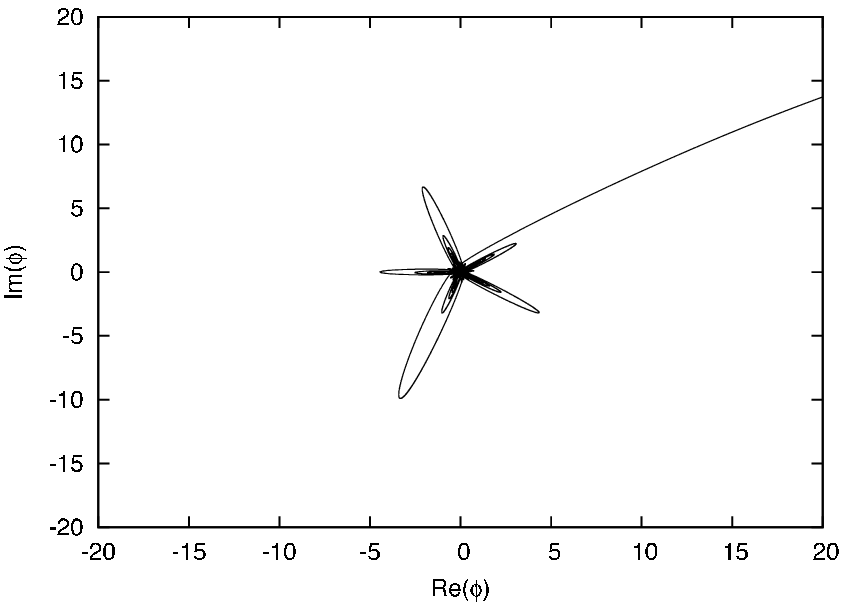}
		\caption{(Upper) Typical motion of the field $\phi$ in the 
		complex plane when $T_R=10^6$GeV. 
		Field value is normalized by $10^{10}$GeV. 
		(Lower) The motion near the origin is magnified}
		\label{fig:Rephi_Imphi_Te5}
	\end{center}
\end{figure}

Thus, for high enough reheating temperature Affleck-Dine 
baryogenesis can work irrespective of the charge-breaking 
minima of the potential.
To confirm this statement, we have performed numerical 
calculation with full scalar potential including 
finite-temperature effect.
For simplicity, we set $M=M_P$ (in next section, we see that 
this choice is valid to obtain a proper amount of baryon 
asymmetry).
As explained above, $T_R\gtrsim 10^6$GeV is needed to 
obtain appropriate motion of the flat direction.
In Fig.~\ref{fig:Rephi_Imphi_Te5}, we show the result 
when $T_R=10^6$GeV and $m_{3/2} = 100$TeV. 
Clearly one can see $\phi$ field falls into the origin 
with angular motion, which indicates that
the Affleck-Dine baryogenesis works. 
The resultant baryon asymmetry is analyzed in the next section.


%

\begin{figure}[htbp]
	\begin{center}
		\includegraphics[width=0.7\linewidth]{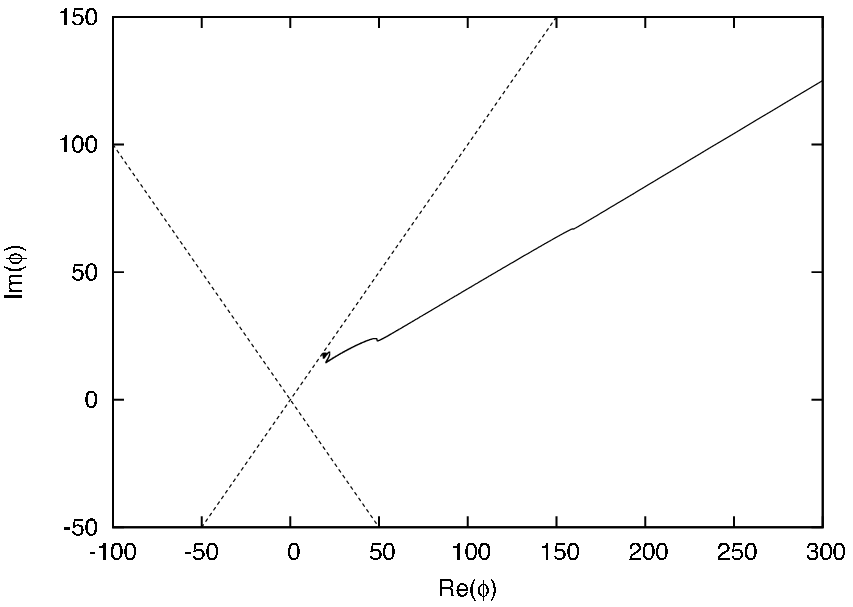}
		\includegraphics[width=0.7\linewidth]{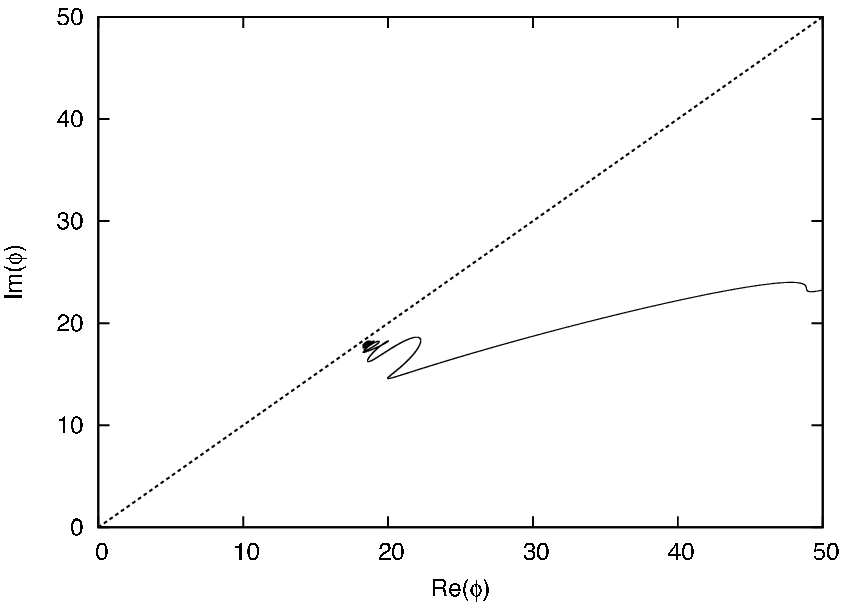}
		\caption{(Upper) Typical motion of the field $\phi$ in the 
		complex plane when $T_R=10^5$GeV. 
		Dotted line represents the valley of the potential from 
		soft A-term. (Lower) The motion near the minimum is magnified}
		\label{fig:Rephi_Imphi_Te4}
	\end{center}
\end{figure}

On the other hand, in Fig.~\ref{fig:Rephi_Imphi_Te4} 
the result for $T_R=10^5$GeV is shown.
In this case, finite-temperature effect is insufficient 
to take the $\phi$ field into the origin,
and finally it is trapped at the charge-breaking displaced 
minimum (\ref{breakmin}).
Therefore, in order to make Affleck-Dine scenario successful 
in anomaly-mediation models,
at least reheating temperature $T_R \gtrsim 10^6$GeV is needed 
though this value varies depending on the cut-off scale $M$.


%


\subsection{Baryon number generation}  

Let us estimate the baryon number created in this process.
Actually the $LH_u$ direction generates lepton number, 
but electroweak 
sphaleron process quickly converts it into baryon 
number~\cite{Kuzmin:1985,Khlebnikov:1988}.
From eq.(\ref{charge}) and equation of motion of $\phi$
\begin{equation}
	\ddot \phi +3H \dot \phi +\frac{\rmd V}{\rmd \phi^*}=0,
\end{equation}
we obtain
\begin{equation}
	\dot n+3Hn=2N \mathrm{Im} 
	\left(\phi \frac{\partial V}{\partial \phi}\right).
\end{equation}
Baryon number per comoving volume is almost fixed at the 
beginning of oscillation.
Integrating above expression, we obtain
\begin{equation}
	n(t_{\mathrm{OS}}) ~\sim ~ N \delta_e m_{3/2}
	(H_{\mathrm{OS}} M^{n-3})^{\frac{2}{n-2}},  
	\label{n(t_OS)}
\end{equation}
where $\delta_e = \sin (\arg a_m + n\arg \phi )$ 
represents the degree of CP violation, which is 
naturally expected to be of order 1. 
Notice that the reheating in which the inflaton
decays completely takes place after the AD field starts
oscillation. After reheating 
the baryon-to-entropy ratio is estimated as
\begin{equation}
	\frac{n}{s}=\frac{n(t_R)}{s(t_R)}\sim 
	N \frac{\delta_e m_{3/2} T_R}{H_{\mathrm{OS}}^2M_P^2}
	(H_{\mathrm{OS}} M^{n-3})^{\frac{2}{n-2}}.  
	\label{n/s}
\end{equation}
In the present $LH_u$ case, $H_{\mathrm{OS}}$ 
is given by eq.~(\ref{Hos}),
\begin{equation}
	\frac{n}{s}\sim 10^{-10} \delta_e 
	\left ( \frac{0.1}{\alpha} \right )
	\left(\frac{m_{3/2}}{50\mathrm{TeV}} \right)
	\left( \frac{10^{-7}\mathrm{eV}}{m_\nu}
	\right)^{\frac{3}{2}}  
	\label{baryonasym}
\end{equation}
where we have used the fact that $m_\nu$ is given by 
\begin{equation}
	m_\nu \sim \frac{\langle H_u \rangle ^2}{M}
	=\frac{\sin ^2 \beta}{M}(174 \mathrm{GeV})^2 
\end{equation}
with $\sin \beta \sim 1$.
Interestingly, the baryon-to-entropy ratio~(\ref{baryonasym}) 
is independent of reheating temperature~\cite{Fujii:2001}.
This is because lowering reheating temperature tends to 
generate small baryon-to-entropy ratio, but
on the other hand, the epoch of oscillation due to 
the thermal logarithmic term becomes late 
and leads to larger baryon number.
As a result, these two effects cancel and baryon-to-entropy 
ratio becomes independent of reheating temperature,
as far as the only requirement $T_R \gtrsim 10^5$GeV 
is satisfied. 
The mass of neutrino should be less than $\sim 10^{-7}$~eV
for successful baryogenesis. 
Note that due to largeness of $m_{3/2}$, constraint 
on the neutrino mass for successful baryogenesis is 
weaker than that of usual gravity-mediation 
case~\cite{Fujii:2001}.
 
\subsection{Some comments} \label{sec:comments}

Some comments are in order.

First, we comment on instability of AD condensate 
and Q-ball formation in our scenario.
If Q-balls are formed, almost all charges are trapped into 
them~\cite{Kasuya:2000} and the subsequent evolution 
becomes complicated.
When finite-temperature effect is neglected,
whether the relevant AD condensate is stable or not is 
determined by quantum correction to the mass squared 
\begin{equation}
	m^2\left\{ 1+K\log \left( \frac{|\phi |^2}{M^2} \right) 
	\right\}.  
	\label{RGmass}
\end{equation}
If $K<0$, instability develops and finally Q-balls are formed.
For $LH_u$ direction, $K$ is positive and Q-balls are not 
formed~\cite{Enqvist:2000}.
But when oscillation of AD condensates occurs due to 
finite-temperature effect, instability can develop and result 
in formation of Q-balls.
However, the resultant charge of Q-balls are so small
that they can not survive until temperature becomes 
lower than the electroweak scale as shown in 
Sec.~\ref{sec:Q-ball}.
Thus, Q-ball formation does not have any non-trivial 
cosmological consequence, and we do not bother to 
worry about complication due to Q-ball formation.

Then, what about $n=4$ flat directions other than $LH_u$?
It is known that other $n=4$ directions in MSSM conserve 
$B-L$~\cite{Gherghetta:1996}. 
Thus sphaleron effects completely wash out the created 
baryon asymmetry.
Although sufficiently large  Q-ball can protect baryon 
asymmetry from sphaleron effect,
such large Q-balls do not seem to be created in the presence 
of early oscillation as described above. 

Finally, we comment on the dark matter candidate in our 
scenario.
Thermal relic of wino LSP in anomaly-mediation models 
can not explain the observed dark matter density due to 
their large annihilation cross section~\cite{Mizuta:1993},
unless it is as heavy as 2 TeV.
Non-thermally produced wino from Q-ball decay
is possible candidate~\cite{Fujii:2002},
but in our model large Q-balls are not formed, 
so this possibility is excluded.
Thus, we can not explain in this model both the baryon 
asymmetry and dark matter simultaneously,
and we need other particle such as axion to account
the present density of matter in the universe.

\section{The case of $n=6$ flat direction }  \label{sec:n=6}

The similar analysis can be applied to $n=6$ flat direction.
There are some flat directions in MSSM which are lifted by 
$n=6$ non-renormalizable superpotentials.
The most interesting direction is $udd$ direction, 
which is responsible for B-ball baryogenesis~\cite{Enqvist:1998}.
In usual gravity-mediation case, 
Q-balls associated with AD condensate corresponding to $n=6$
flat direction can survive below the freeze-out temperature
of LSP, and subsequent decay produce baryon number 
and non-thermal LSP.
In some models, this can naturally explain both the baryon 
asymmetry and dark matter~\cite{Fujii&Yanagida:2002,Fujii:2002}.

However, in our model, to avoid charge or color breaking minima, 
early oscillation due to the thermal logarithmic term is needed.
Since the $udd$ direction is expected to have a negative 
coefficient for thermal logarithmic term ($a < 0$ in 
eq.~(\ref{eq:thermal_log})), early oscillation unlikely occurs.
Thus, the argument similar to $LH_u$ case can not be applied 
to this direction.
But for other $n=6$ direction, e.g. $LLe$ direction, 
it may be possible that AD baryogenesis in anomaly-mediation 
models works.
In this section, we investigate this possibility.

\subsection{Dynamics of $n=6$ flat direction }

First, we study the condition for early oscillation to occur.
Using $|\phi |\sim (HM^3)^{1/4}$, we obtain 
\begin{eqnarray}
	T_R  ~\gtrsim ~ \frac{1}{\alpha} 
	\left( \frac{m_\phi^3 M^3}{M_P^2} \right)^{\frac{1}{4}}  
	 \sim  3\times 10^5 \mathrm{GeV}
	\left( \frac{m_\phi}{100\mathrm{GeV}} \right)^{\frac{3}{4}}
	\left( \frac{M}{10^{15}\mathrm{GeV}} \right)^{\frac{3}{4}}.
\end{eqnarray}
Compared with $n=4$ flat direction, 
higher reheating temperature or smaller cut-off scale are needed.

Next, in order to avoid the unphysical minima, the 
thermal log must hide the valley of the zero-temperature 
potential.
From eq.(\ref{potmin}) with $n=6$,
\begin{equation}
	V(|\phi|_{\mathrm{min}}) \simeq 
	-\frac{2}{75\sqrt 5}m_{3/2}^{\frac{5}{2}}M^{\frac{3}{2}}.
\end{equation}
We require $\alpha^2 T^4 \gtrsim |V(|\phi|_{\mathrm{min}}) |$ 
at the beginning of oscillation.
This leads to
\begin{eqnarray}
	T_R  & \gtrsim & \alpha ^{-1} 
	M^{\frac{3}{4}}M_P^{-\frac{1}{2}}m_{3/2}^{\frac{3}{4}}
	\nonumber \\ 
    &\sim &  8\times 10^5 \mathrm{GeV}
	\left( \frac{0.1}{\alpha}\right) 
	\left( \frac{M}{10^{15}\mathrm{GeV}}\right)^{\frac{3}{4}}
	\left( \frac{m_{3/2}}{10\mathrm{TeV}}\right)^{\frac{3}{4}}.
	\label{constraint:n=6}
\end{eqnarray}
Thus, $T_R \gtrsim 10^6$~GeV is necessary 
for $M\sim 10^{15}$GeV and $m_{3/2} \sim 10$TeV.
When this constraint on the reheating temperature is satisfied,
the dynamics of flat direction is similar to that of 
$LH_u$ direction studied in the previous section.
Until $H$ becomes lower than $H_{\mathrm{OS}}$, 
the radial component of AD field is trapped by the
instant minimum determined by the negative Hubble mass 
term and the non-renormalizable term.
Then, AD field begins to oscillate around their origin, 
and receives angular kick.
The baryon-to-entropy ratio is almost fixed at this epoch. 
Although this reheating temperature seems rather high,
gravitino mass is 
heavy enough to decay before the BBN epoch in anomaly-mediation models.
So high reheating temperature is not a problem. 
We have checked numerically that for $T_R\gtrsim 10^6$GeV, 
$\phi$ rolls down to the origin without trapped 
by the displaced minima.

\subsection{Baryon number generation}

If the constraint on the reheating temperature 
(\ref{constraint:n=6}) is satisfied,
Affleck-Dine mechanism can work with no more difficulty.
From eq.(\ref{n/s}),
\begin{equation}
	\frac{n}{s}\sim \frac{1}{9}
	\delta_e m_{3/2}T_R M_P^{-2} 
	\left( \frac{M}{H_O} \right)^{\frac{3}{2}}  
	\label{n/s(n=6)}
\end{equation}
where $H_{\mathrm{OS}}$ is given by
\begin{equation}
	H_{\mathrm{OS}}\sim 
	(a\alpha^2 T_R^2M_P M^{-\frac{3}{2}})^{\frac{2}{3}}  
	\label{Hos(n=6)}
\end{equation}
from eq.~(\ref{osc.condition}).
Substituting this into eq.~(\ref{n/s(n=6)}), we obtain
\begin{eqnarray}
	\frac{n}{s}& \sim & 
	\frac{\delta_e m_{3/2}M^3}{9\alpha^2 T_R M_P^3} \\
	& \sim & 10^{-10} \delta_e
	\left ( \frac{0.1}{\alpha} \right )^2
	\left( \frac{m_{3/2}}{10\mathrm{TeV}} \right)
	\left(\frac{10^8\mathrm{GeV}}{T_R} \right)
	\left(\frac{M}{10^{16}\mathrm{GeV}} \right)^3.  
	\label{n/s:n=6}
\end{eqnarray}
Thus, we can obtain a proper amount of baryon asymmetry 
with parameters consistent with the 
constraint~(\ref{constraint:n=6}).

\section{Q-ball formation}  \label{sec:Q-ball}

In the previous section, we have showed that the
Affleck-Dine mechanism can create a sizable baryon 
asymmetry in anomaly-mediation models
with rather high reheating temperature.
But it is non-trivial matter whether the created 
baryon asymmetry actually provides the baryon density 
of the universe which is required by BBN or CMB anisotropy.
This is because Q-balls may be formed 
and almost all baryon charge are trapped into them.
If charge of Q-balls is large and they are stable, 
the evolution of the AD field and resultant baryon 
asymmetry of the universe may be changed significantly.
In this section, we see that Q-ball formation in our 
models have no great importance on cosmology.

Generally, perturbations to the AD fields $\phi$ grow 
when their potential is less steep than $\phi^2$ 
due to the negative pressure.
In our models, oscillation of the AD fields is controlled 
by thermal logarithmic term,
and hence the instability develops into formation of Q-balls.
Since this is similar to gauge-mediation type Q-ball,
the same analysis as gauge-mediation case can be applied.

First note that although the whole dynamics of Q-ball 
formation is highly non-linear, the radius of Q-balls 
is determined by the fastest growing mode in perturbative 
analysis.
This is checked by numerical calculation based on 
lattice simulation~\cite{Kasuya:2000}.
Roughly speaking, the wave length of the most growing 
mode is comparable to the Hubble radius at the epoch of 
oscillation of AD fields, 
\begin{equation}
	\frac{|\bold{k}|}{a}\sim H_{\mathrm{OS}}\sim 
	\frac{T_{\mathrm{OS}}^2}{|\phi(t_{\mathrm{OS}})|},
\end{equation}
where $T_{\mathrm{OS}}$ denotes the temperature at 
the beginning of oscillation and is given by 
$T_{\mathrm{OS}}\sim (T_R^2 H_{\mathrm{OS}} M_P)^{1/4}$.
Thus, we expect early oscillation due to thermal effect 
tends to yield smaller Q-balls.
The resultant charge inside Q-balls $Q$ is given by 
$Q\sim H_{\mathrm{OS}}^{-3}n_B(t_{\mathrm{OS}})$.
Here we note that eq.(\ref{n(t_OS)}) can be written 
in the convenient form 
$n_B(t_{\mathrm{OS}}) \sim \epsilon H_{\mathrm{OS}} 
|\phi(t_{\mathrm{OS}})|^2$,
where $\epsilon$ denotes ellipticity of the orbit of 
AD field, $\epsilon=m_{3/2}/H_{\mathrm{OS}}$.
The result is,
\begin{equation}
	Q\sim \epsilon 
	\frac{|\phi(t_{\mathrm{OS}})|^4}{T_{\mathrm{OS}}^4} 
	\sim \epsilon 
	\left(\frac{M}{T_{\mathrm{OS}}}\right)^{4\frac{n-3}{n-1}}.  
	\label{Qcharge}
\end{equation}
In fact, Q-ball formation begins slightly later than the 
oscillation of AD field and the number of Q-balls per 
Hubble horizon is expected to be more than one.
In ref.~\cite{Kasuya:2001}, it is found that the maximum 
charge of Q-balls is fitted by the formula
when the ellipticity $\epsilon $ is much smaller than 1,
\begin{equation}
	Q_{\mathrm{max}}=\beta 
	\left(\frac{\phi(t_{\mathrm{OS}})}{T_{\mathrm{OS}}}\right)^4 
	\label{Qcharge2}
\end{equation}
with $\beta \sim 6\times 10^{-4}$ from lattice simulation.
This agrees with eq.~(\ref{Qcharge}) except for numerical 
factor which can not be determined analytically. 
Eq.~(\ref{Qcharge2}) is independent of $\epsilon$ because
anti-Q-balls are also produced so that the net baryon number 
is small.  
In the case of early oscillation, since it is expected that 
$\epsilon \ll 1$, 
we use eq.~(\ref{Qcharge2}) in the following.
Now we estimate the charges of Q-balls in $n=4$ and $n=6$ case, respectively.

\subsection{$n=4$ case}

From eq.~(\ref{Hos}), the temperature at the oscillation 
$T_{\mathrm{OS}}$ is given by 
\begin{equation}
	T_{\mathrm{OS}}\sim 
	(\alpha T_R^3 M_P^{\frac{3}{2}}M^{-\frac{1}{2}})^{\frac{1}{4}}.
\end{equation}
Thus, the charge of Q-balls is estimated as
\begin{eqnarray}
	Q & \sim & \beta 
	 \left( \frac{M}{T_{\mathrm{OS}}} \right)^{\frac{4}{3}} 
	 \nonumber \\
	& \sim  & 
	3\times 10^{9} 
	\left(\frac{0.1}{\alpha} \right)^{\frac{1}{3}}
	\left(\frac{\beta}{6\times10^{-4}} \right)
	\left(\frac{10^6\mathrm{GeV}}{T_R} \right)
	\left(\frac{M}{M_P} \right)^{\frac{3}{2}}.		
\end{eqnarray}
With the aid of eq.~(\ref{n/s}), this can be rewritten as
\begin{eqnarray}
	Q   & \sim & 1\times 10^9
	~\delta_e^{-1}
	\left(\frac{\alpha}{0.1} \right)^{\frac{2}{3}}
	\left(\frac{\beta}{6\times10^{-4}} \right) 
	\nonumber \\
	&  & \times  
	\left(\frac{10^6 \mathrm{GeV}}{T_R} \right)
	\left(\frac{50\mathrm{TeV}}{m_{3/2}} \right)
	\left(\frac{n_B/s}{10^{-10}} \right).   
	\label{Q:n=4}
\end{eqnarray}
As shown in~\ref{Appendix}, in order to survive  evaporation 
in high temperature plasma, $Q\gtrsim 10^{18}$ is needed.
Thus, even if Q-balls are formed, they are expected to 
evaporate completely well above $T\sim 100$GeV and 
the estimation of baryon asymmetry in Sec.~\ref{sec:n=4} 
need not be changed.

\subsection{$n=6$ case}

In this case, $H_{\mathrm{OS}}$ is given by 
eq.~(\ref{Hos(n=6)}).  Then, the temperature 
at the oscillation is 
\begin{equation}
	T_{\mathrm{OS}} \sim \alpha^{\frac{1}{3}} 
	T_R^{\frac{5}{6}} M^{-\frac{1}{4}}M_P^{\frac{5}{12}}.
\end{equation}
The charge of Q-balls in the $n=6$ case becomes
\begin{eqnarray}
	Q & \sim & \beta 
	\left(\frac{M}{T_{\mathrm{OS}}}\right)^{\frac{12}{5}} 
	\nonumber \\
	& \sim & 2\times 10^{10} 
	\left(\frac{0.1}{\alpha} \right)^{\frac{4}{5}}
	\left(\frac{\beta}{6\times10^{-4}} \right)
	\left(\frac{10^7\mathrm{GeV}}{T_R} \right)^{2}
	\left(\frac{M}{10^{15}\mathrm{GeV}} \right)^3.
\end{eqnarray}
Using eq.~(\ref{n/s:n=6}), this is rewritten  
in terms of $n/s$ as
\begin{eqnarray}
	Q & \sim & 2\times 10^{12}
	~\delta_e^{-1}
	\left(\frac{\alpha}{0.1} \right)^{\frac{6}{5}}
	\left(\frac{\beta}{6\times10^{-4}} \right) 
	\nonumber \\
	&  & \times \left(\frac{10^7 \mathrm{GeV}}{T_R} \right)
	\left(\frac{10\mathrm{TeV}}{m_{3/2}} \right)
	\left(\frac{n_B/s}{10^{-10}} \right).   
	\label{Q:n=6}
\end{eqnarray}
This is also so small that Q-balls can not 
survive evaporation.

\section{Conclusion}  \label{sec:conclusion}

We have investigated the Affleck-Dine mechanism 
in anomaly-mediated SUSY breaking models.
In contrast to previous studies,
we have found that early oscillation due to 
finite-temperature effects can drive flat directions 
into the correct vacuum and a proper amount of baryon 
asymmetry can be generated for neutrino 
mass about $m_\nu \lesssim 10^{-7}$~eV in the case of $LH_u$ direction.
Our scenario requires somewhat high reheating temperature,
but this leads to no cosmological difficulties such as gravitino 
problem since  gravitino is heavy enough and decay 
before the onset of BBN in anomaly-mediation models.

We have also investigated the same mechanism for $n=6$ 
flat direction case.
It is found that for natural range of parameters,
proper amount of baryon asymmetry can be obtained.
Furthermore, we also have discussed consequences of 
Q-ball formation. It is shown that all Q-balls evaporate 
in high-temperature plasma. Therefore, the Q-ball formation
does not complicate the baryogenesis process. 

\appendix
\section{Evaporation of Q-balls} \label{Appendix}
 
A Q-ball is a non-topological soliton whose stability is 
ensured by global $U(1)$ symmetry.
But it can release its charge in some manner.
Here we concentrate on the following two process.
The first is decay of AD field into pair of fermions or 
lighter bosons, and the second is evaporation and diffusion 
effects in thermal bath.
In this Appendix, we give a rough estimation of the total 
amount of evaporated charge from Q-balls.

\subsection{Decay into light particles}

It is known that the energy per charge inside Q-balls 
of gauge-mediation type 
is proportional to $Q^{-1/4}$~\cite{Kusenko:1998}. 
Thus, for large enough $Q$, Q-balls are stable against 
decay into nucleons.
The energy per unit charge of gauge-mediated type 
Q-ball is given by 
\begin{equation}
	\frac{E_Q}{Q}\sim T Q^{-\frac{1}{4}}.
\end{equation}
This leads to the stability condition,
\begin{equation}
	Q > \left(\frac{T}{m_N} \right)^4
\end{equation}
where $m_N$ denotes the mass of nucleon, $m_N\sim1$GeV.
Thus, for $T\lesssim Q^{1/4}$GeV, Q-balls become stable 
against decay into nucleons, although decay into light 
neutrinos is possible for leptonic Q-balls (L-balls).
But as temperature becomes low,
as explained in the next subsection,
gauge-mediated type Q-balls dominated by thermal 
logarithmic potential are deformed or converted into 
gravity-mediated type.
Even if Q-balls survive from evaporation, 
eventually they decay into free fermions or bosons
since gravity-mediated type Q-balls have energy-to-charge 
ratio comparable to $m_\phi$, which is much larger 
than $m_N$.

Thus, we consider the decay process of Q-balls into 
fermion pair.
This occurs only from the surface of Q-ball since 
Pauli exclusion principle forbids the decay into fermions 
inside Q-ball~\cite{Cohen:1986}.
This sets upper bound on the decay rate of Q-balls which 
can easily be saturated,
\begin{equation}
	\left(\frac{\rmd Q}{\rmd t} \right)_{\mathrm{fermion}} 
	\leq \frac{A\omega^3}{192\pi^2},
\end{equation}
where $A$ denotes the surface area of Q-balls. 
Decay into pair of bosons is possible and may be largely enhanced compared with 
the case of fermions 
if there exists lighter scalar fields than AD fields.
However, since this bosons become heavy inside Q-balls, 
the decay into lighter bosons only occurs through loop 
diagrams suppressed by the large effective 
mass~\cite{Enqvist:1998}, which leads to 
the enhancement factor $f_s\lesssim 10^3$, defined by
\begin{equation}
	\left(\frac{\rmd Q}{\rmd t} \right)_{\mathrm{boson}}
	=f_s 
	\left(\frac{\rmd Q}{\rmd t} \right)_{\mathrm{fermion}}.
\end{equation}
Using $A\sim 4\pi R_Q^2 \sim 4\pi |K|^{-1}m_\phi^{-2}$ 
for gravity-mediated type Q-balls,
we obtain the decay temperature of Q-balls
\begin{equation}
	T_d \sim 18\mathrm{GeV}\sqrt {f_s}
	\left(\frac{0.01}{|K|} \right)^{\frac{1}{2}} 
	\left(\frac{m_\phi}{100\mathrm{GeV}} \right)^{\frac{1}{2}}
	\left(\frac{10^{18}}{Q} \right)^{\frac{1}{2}}.
\end{equation}
For sufficiently large $Q$, this can become lower than 
the freeze-out temperature of dark matter, 
$T_f\sim m_{\mathrm{DM}}/20$.
If this is the case, wino dark matter, which is the natural 
consequence of anomaly mediation model, produced by 
Q-ball decay can amount to desired abundance of dark 
matter~\cite{Fujii:2002}.
But it should be noticed that $udd$ direction is invalid 
because the coefficient of thermal logarithmic term is 
expected to be negative and hence the early oscillation 
does not occur.
Furthermore, if one wants to explain baryon asymmetry 
and dark matter in this scenario,
$LLe$ or other pure leptonic direction does not work either,
since created lepton number is protected from sphaleron 
effect inside the Q-ball until they decay at temperature 
below the electroweak scale.
Other flat directions lifted by $n=6$ superpotential is 
attractive candidates~\cite{Gherghetta:1996},
but to obtain large $Q$ is difficult in our scenario
(see eqs.(\ref{Q:n=4}) and (\ref{Q:n=6})).

\subsection{Evaporation and diffusion}

Q-ball formation is non-adiabatic process, and almost 
all charges are trapped into the Q-balls.
This configuration is energetically stable, 
but in finite temperature environment, this is not 
always the case.
In thermal bath there can exist free particles 
carrying charge surrounding Q-balls. 
The minimum of free energy is achieved when all charges 
are distributed in the form of Q-plasma.
However, in actual situation, the evaporation of charge 
from Q-balls are not so sufficient in cosmic time scale.
Thus, the mixture of plasma state and Q-ball state is 
realized.
Then, it is important to know that how and what 
amount of charge of Q-balls is released into outer region.

Q-balls emit their charge through two process, 
evapolation~\cite{Laine:1998}
and diffusion~\cite{Banerjee:2000}.
Generally, as we see below, at high temperature the latter 
determines the emission rate of charge from Q-balls.
First we estimate the evaporation rate from Q-balls.
This occurs when the value of chemical potential of 
Q-balls ($\mu_Q$) and surrounding plasma ($\mu_p$) 
differs significantly.
The evaporation rate is
\begin{equation}
	\Gamma_{\mathrm{evap}}=-4\pi R_Q^2 \xi 
	(\mu_Q-\mu_p)T^2 , 
\end{equation}
where $\xi$ is given by
\begin{equation}
	\xi = \left \{ 
	   \begin{array}{ll}
	   1 ~~~~~&(T>m_\phi) \\
	   \left(\frac{T}{m_\phi} \right)^2 ~~~~~&(T<m_\phi).
            \end{array}
            \right.
\end{equation}
But in fact around the edge of Q-balls, 
chemical equilibrium between plasma and 
Q-matter are achieved and charge inside the
Q ball cannot come out at high temperature. 
Therefore, the charge in the
`atmosphere' of the Q ball should be taken away 
by diffusion in order for further charge evaporation.
In this situation, charge transfer from inside Q-balls 
into plasma are determined by diffusion effect 
rather than above evaporation rate,
\begin{equation}
	\Gamma_{\mathrm{diff}}=-4\pi DR_Q\mu_Q T^2 
	\sim -4\pi aT.
\end{equation}
where we have used $D=a/T$ with $a=4-6$, 
$\mu_Q\sim T Q^{-1/4}$ and $R_Q \sim T^{-1}Q^{1/4}$.
when $T>m_\phi$, 
$\Gamma_{\mathrm{diff}}<\Gamma_{\mathrm{evap}}$ and 
the charge transfer is controlled by diffusion effects.

In our model, the thermal logarithmic potential 
dominates when AD field oscillates and Q-ball formation 
takes place.
However, as temperature becomes low, the thermal effect 
ceases to be the dominant component of  of the potential.
As a rough estimation, this occurs when 
$T^4 \sim m_\phi^2 |\phi_{\mathrm{hill}}|^2$,
that is, $T\sim T_c \sim 10^6$GeV.
If $|\phi |\ll |\phi_{\mathrm{hill}}|$ at this epoch,
soft mass term determines the properties of Q-balls.
If $K$ in (\ref{RGmass}) is negative, the Q-ball configuration 
is a gravity-mediation type,
$R_Q\sim m_\phi |K|^{-1/2}, \mu_Q\sim m_\phi$.
Otherwise, Q-ball configuration does not stable any more 
and will collapse.
In this case, Q-ball formation is irrelevant to baryogenesis.
In the following, we consider the possibility that
at $T<T_c$, the configuration of Q-balls is changed 
into gravity-mediation type.
In our model, the reheating temperature $T_R$ must be rather 
high as shown in previous sections.
Thus, in the following, we assume 
$T_{\mathrm{OS}}>T_R>T_c>m_\phi$.
Then, we obtain
\begin{equation}
	\frac{\rmd Q}{\rmd T}=\left \{
	\begin{array}{ll}
	\frac{32\pi aT_R^2 M_P}{3T^4} ~~~~~&(T>T_R) \\[0.5em]
	\frac{16\pi aM_P}{T^2} ~~~~~&(T_R>T>T_c) \\[0.5em]
	\frac{16\pi aM_P}{|K|^{1/2} T^2} ~~~~~&(T_c>T>T_*) \\[0.5em]
	\frac{16\pi M_P}{|K| m_\phi T} ~~~~~&(T_*>T>m_\phi)\\[0.5em]
	\frac{16\pi M_P T}{|K| m_\phi^3 } ~~~~~&(T<m_\phi),
	\end{array}
	\right.
\end{equation}
where $T_*$ is defined by $T_*=a|K|^{1/2}m_\phi$.
Although we have assumed $T_*>m_\phi$, this assumption 
does not much affect the following analysis.
Now let us estimate the total amount of the evaporated 
charge $\Delta Q$ and examine whether or not Q-balls  
can survive in our model.
Integrating above evaporation or diffusion rate, 
we obtain
\begin{eqnarray}
	\Delta Q(T>T_R) &\sim &\frac{32\pi aM_P}{9T_R},\\
	\Delta Q(T_R>T>T_c)&\sim &\frac{16\pi aM_P}{T_c},\\
	\Delta Q(T_c>T>T_*)&\sim &\frac{16\pi aM_P |K|^{-1/2}}{T_*},\\
	\Delta Q(T_*>T>m_\phi)&\sim &\frac{16\pi M_P|K|^{-1}}{m_\phi},\\
	\Delta Q(T<m_\phi)&\sim &\frac{8\pi M_P|K|^{-1}}{m_\phi},
\end{eqnarray}
which are estimated as
\begin{eqnarray}
	\Delta Q(T>T_R) &\sim & 1\times 10^{13}
	\left(\frac{10^7\mathrm{GeV}}{T_R} \right),\\
	\Delta Q(T_R>T>T_c)& \sim & 2\times 10^{14} 
	\left(\frac{10^6\mathrm{GeV}}{T_c} \right),\\
	\Delta Q(T_c>T>T_*)&\sim & 2\times 10^{17} |K|^{-\frac{1}{2}}
	\left(\frac{10^3\mathrm{GeV}}{T_*} \right),\\
	\Delta Q(T_*>T>m_\phi)& \sim & 2\times 10^{17} |K|^{-1} 
	\left(\frac{10^3\mathrm{GeV}}{m_\phi} \right),\\
	\Delta Q(T<m_\phi) & \sim & 1\times 10^{17} |K|^{-1} 
	\left(\frac{10^3\mathrm{GeV}}{m_\phi} \right).
\end{eqnarray}
From these, in order for Q-balls to survive evaporation,
\begin{equation}
	Q \gtrsim 10^{18}
\end{equation}
is required.
Therefore, the charge transfer is enough to evaporate 
all charge from Q-balls for both $n=4$ and $n=6$ cases.
Even if Q-balls are stable against decay into lighter 
particles, diffusion and evaporation effects can 
sufficiently transfer their charge into outer environment.
Since our model requires high reheating temperature,
this effect is unavoidable.
After all, Q-balls can not survive until temperature 
becomes lower than about 100~GeV,
where electroweak phase transition occurs. 


\end{document}